\documentclass[aps,pre,reprint,superscriptaddress]{revtex4-1}
\usepackage{graphicx}
\usepackage{amsmath}
\usepackage{amssymb}
\usepackage{bm}
\usepackage{physics}
\usepackage{subfigure}
\usepackage[colorlinks=true,linkcolor=blue,urlcolor=blue,citecolor=blue]{hyperref}
\usepackage{times}

\begin{document}

\title{Dissipative features of the driven spin-fermion system}
\author{Ruofan Chen}
\affiliation{Science and Technology on Surface Physics and Chemistry Laboratory, Mianyang 621908, China}
\author{Xiansong Xu}
\affiliation{Science and Math Cluster, Singapore University of Technology and Design, 8 Somapah Road, Singapore 487372}
\date{\today}

\begin{abstract}
  We study a generic spin-fermion model, where a two-level system
  (spin) is coupled to two metallic leads with different chemical
  potentials, in the presence of monochromatic driving fields. The
  real-time dynamics of the system is simulated beyond the Markovian
  limit by an iterative numerically exact influence functional path
  integral method.  Our results show that although both system-bath
  coupling and chemical potential difference contribute to
  dissipation, their effects are distinct. In particular, under
  certain drivings the asymptotic Floquet states of the system exhibit
  robustness against a range of system-bath coupling strength: the
  asymptotic behaviors of the system are insensitive to different
  system-bath coupling strength, while they are highly tunable by the
  chemical potential difference of baths. Further simulations show
  that such robustness may be essentially a result of the interplay
  between driving, bath electronic structure and system-bath
  coupling. Therefore the robustness could break down depending on the
  characteristics of the interplay. In addition, under fast linearly
  polarized driving the quantum stochastic resonance is demonstrated
  that stronger system-bath coupling (stronger dissipation) enhances
  rather than suppresses the amplitude of coherent oscillations of the
  system.
\end{abstract}
\maketitle

\section{Introduction}
A wide range of physical and chemical systems can be effectively
described by quantum two-level systems (TLSs). The most common example
for a TLS is a particle of total spin \(\frac{1}{2}\) under an
external magnetic field which is shown in nuclear magnetic resonance
experiments \cite{schmidt1992-floquet}. Another common situation is a
particle moving in an effective double-well potential in which only
the lowest energy doublet is occupied
\cite{leggett1987-dynamics,farrelly1993-two}. A well-known example is
the ammonia molecule, NH\(_{\text{3}}\): quantum mechanically the
hydrogens can tunnel back and forth between two potential minima
\cite{hund1927-on,rohrer1988-the}. The TLS is also the simplest
nontrivial physical model used as a starting point to study
time-dependent quantum systems. The explicitly time-dependent quantum
problem generates a variety of novel phenomena that are not accessible
within stationary quantum mechanics. A comprehensive review is given
by Grifoni and H\"{a}nggi \cite{grifoni1998-driven}.

An isolated TLS is an ideal model and often fails to describe thermal
and dynamical properties of real physical or chemical systems when the
system is in contact with external environments. If environments can
be effectively described as a collection of harmonic oscillators, then
we obtain the so-called spin-boson model
\cite{chang1985-dissipative,leggett1987-dynamics} which has been
widely studied and exhibits rich phenomena
\cite{grifoni1998-driven,weiss1993-quantum}. The environments can also
be fermionic and in this case we have the spin-fermion model
\cite{chang1985-dissipative,segal2007-nonequilibrium,mann2016-dissipative}. The
spin-boson and spin-fermion models represent the simplest nontrivial
quantum dissipative models and are related to various physical and
chemical problems. They are relevant for modeling charge transfer in
photosynthesis \cite{gilmore2005-spin}, the Kondo problem for magnetic
impurities \cite{kondo1964-resistance,anderson1961-localized}, quantum
stochastic resonance
\cite{gammaitoni1998-stochasticresonance,makarov1995-stochastic,wagner2019-quantum}
and quantum decoherence in the context of a superconducting charge
qubit
\cite{makhlin2001-quantum,paladino2002-decoherence,grishin2005-low,sousa2005-ohmic}.

The possibility of controlling the time evolution of the molecular
system by time-dependent external field has long appealed to chemists
and physicists in order to lead a chemical reaction toward the desired
product, design better nanodevices, etc. For instance, it is shown
that time-dependent control can boost the thermoelectric efficiency of
nanodevices \cite{zhou2015-boosting}. Manipulating the time evolution
of a molecular system requires controlling dissipative
mechanisms. Understanding such mechanisms is of both practical and
fundamental theoretical interest. A driven spin-boson or spin-fermion
model could serve as a simple but nontrivial model for theoretical
investigations.

While the spin-boson model has been extensively studied both in the
presence
\cite{grifoni1995-cooperative,weiss1993-quantum,grifoni1998-driven,makarov1995-control,makri1997-universal,makri1997-stabilization,hartmann1998-dissipative,hartmann2000-driven,shuang2000-cooperativity,thorwart2000-iterative,hausinger2010-dissipative,magazzu2018-asymptotic}
and absence
\cite{leggett1987-dynamics,weiss1993-quantum,chen1989-degenerate,makarov1994-path,makri1995-numerical,salkola1996-coupled,makri1999-iterative,golosov1999-efficient,mitra2005-spin,segal2005-spin,allahverdyan2005-work,nesi2007-spin,zhou2008-solving,porras2008-mesoscopic,segal2014-heat,shapourian2016-dynamical,wall2016-simulating}
of external driving, the spin-fermion model is less well
understood. In particular, the interplay of external driving on the
system and the system-environment coupling strength is less explored
due to the limitations of both theoretical and numerical tools. In
this article we employ an iterative numerically exact influence
functional path integral method to study the driven spin-fermion
model. This method was first developed by Makarov and Makri and
applied to the spin-boson model without driving
\cite{makarov1994-path,makri1995-numerical}, then it was successfully
applied to investigate the driven spin-boson model
\cite{makri1997-universal,makarov1995-control,makri1997-stabilization}. It
is a nonperturbative method beyond the Markovian limit and thus well
suited for handling real-time dynamics problems. Later Segal \emph{et
  al.}  adopted a more flexible discretized scheme for tracing out the
bath and generalized this method to investigate the spin-fermion and
some other generic models in the absence of driving
\cite{segal2010-numerically,segal2011-nonequilibrium,simine2013-path,segal2013-qubit,agarwalla2017-anderson}.
In this article we adopt such a discretized scheme and extend the
method to the spin-fermion model in the presence of driving.  Our
implementation is compared with the Born-Markov master equation in the
Floquet basis \cite{hone2009-statistical}. In Sec. \ref{sec:method} we
give a brief review of the method and the details of the model are
given in Sec. \ref{sec:model}.

In this article we investigate the spin-fermion model under
monochromatic driving at zero temperature with two fermionic leads
kept at different chemical potentials. Monochromatic circularly and
linearly polarized driving fields are considered in
Sec. \ref{sec:results}.

A noticeable phenomenon appears that under certain drivings the
asymptotic Floquet states of the system exhibit robustness against
different system-bath coupling strength: the asymptotic behaviors of
the system tend to be almost the same even with different coupling
strength. It seems that although both the system-bath coupling and the
chemical potential difference of the baths affect the dissipative rate
greatly, the asymptotic behavior of the system is insensitive to
system-bath coupling strength but dominated by the chemical potential
difference of baths.  Such a feature may be useful in designing a
nanodetector driven by an external field to detect the electronic
structure of the environment. In this case the driving field must be
also considered as a part of the detector and the time-dependent form
of the driving field should be treated as a parameter of essential
importance.

More simulations indicate that there may exist a complex interplay
between driving, bath electronic structure and system-bath
coupling. Such robustness may be essentially a result of the interplay
and can break down depending on the characteristics of the interplay.
Moreover, under fast linearly polarized driving (in
Sec. \ref{sec:linear-z}), quantum stochastic resonance is shown that
stronger system-bath coupling (stronger dissipation) enhances rather
than suppresses the amplitude of coherent oscillations of the system.

Moreover, a convergence test is given in Appendix
\ref{sec:convergence} and a benchmark against the Born-Markov master
equation in the Floquet basis is given in Appendix
\ref{sec:floquet-me}.

\section{General Formulation of Iterative Path Integral Method}
\label{sec:method}

Here we give a brief review of the general formalism of the iterative
path integral method (for more details refer to
Refs. \cite{makarov1994-path,makri1995-numerical,segal2010-numerically}).
Let us consider a generic many-body system which is modeled by a
finite system of interest coupled with two noninteracting baths. Let
\(H(t)\) denote the total Hamiltonian and \(\rho(t)\) denote the total
density matrix. Then the time evolution of \(\rho(t)\) is given by
\begin{equation}
     \rho(t)=U(t)\rho(0)U^{\dag}(t),
\end{equation}
where
\begin{equation}
\label{eq:U-operator}
U(t)=\mathrm{T}\exp\qty[-i\int_0^tH(\tau)\dd{\tau}]=
\prod_{t_i=0}^t e^{-iH(t_i)\delta t}.
\end{equation}
Here \(\mathrm{T}\) denotes the chronological ordering symbol and the
product is understood in that we take the limit over all the
infinitesimal intervals \(\delta t\) between zero and
\(t\). Basically, the evolution is split into \(N\) pieces for which
\(\delta t=t/N\) with \(N\to\infty\). Now we introduce the reduced
density matrix of the system, \(\rho_S=\Tr_B\rho\), which is obtained
by tracing the total density matrix over the bath degrees of
freedom. The time evolution of \(\rho_S(t)\) is then exactly given by
\begin{equation}
     \rho_S(s'',s';t)=\Tr_B\mel{s''}{U(t)\rho(0)U^{\dag}(t)}{s'}.
\end{equation}
Employing finite \(\delta t\) in Eq. \eqref{eq:U-operator}
approximates the evolution operator \(U(t)\) into a product of finite
\(N\) exponentials where
\(U(t)\approx\prod_{t_i=0}^t[e^{-iH(t_i)\delta t}]\). Defining the
discrete time evolution operator \(T=e^{iH(t_i)\delta t}\), then the
reduced density matrix can be written as
\begin{equation}
     \rho_S(s'',s';t)=\Tr_B\mel{s''}{(T^{\dagger})^N\rho(0)T^N}{s'}.
\end{equation}
Inserting the identity operator \(\int\dd{s}\ketbra{s}{s}\) between
every two \(T\) and relabeling \(s'',s'\) as \(s_N^+,s_N^-\) yields
\begin{equation}
\label{eq:path-integral}
\begin{split}
     \rho_S(s_N^+,s_N^-;t)&=\int\dd{s}_0^+\cdots\dd{s}_{N-1}^+
       \int\dd{s}_0^-\cdots\dd{s}_{N-1}^-\\
       & \quad\times\Tr_B{}[\mel{s_N^+}{T^{\dag}}{s_{N-1}^+}
	 \mel{s_{N-1}^+}{T^{\dag}}{s_{N-2}^+}\\
       & \quad\times\cdots\mel{s_0^+}{\rho(0)}{s_0^-}\cdots\\
       & \quad\times\mel{s_{N-2}^-}{T}{s_{N-1}^-}
	 \mel{s_{N-1}^-}{T}{s_N^-}].
\end{split}
\end{equation}

The integrand in Eq. \eqref{eq:path-integral} is referred to as the
``influence functional'' \cite{segal2010-numerically} (IF) and denoted
by \(I(s_0^{\pm},\ldots,s_N^{\pm})\). The IF contains the information
of the system and bath degrees of freedom with system-bath
interactions. The IF has an important property that allows us to
greatly simplify the calculation: nonlocal correlations contained in
the IF decay exponentially under certain conditions
\cite{makarov1994-path}, which enables a (controlled) truncation of
the IF. It means in practical calculation we need to only keep a
finite memory length. Basically, for a system under a chemical
potential bias \(\Delta\mu\) at zero temperature the exponentially
decaying of the correlations is guaranteed by finite \(\Delta\mu\),
while in a large-temperature situation (\(T>\Delta\mu\)) the
temperature sets the scale of the memory length that needs to be kept
\cite{weiss2008-iterative,segal2010-numerically}.  Based on this
feature, an iterative scheme for evaluating the path integral has been
developed \cite{makarov1994-path,makri1995-numerical}. The original
quasiadiabatic path integral algorithm was based on the analytical
pairwise form of the IF specific to harmonic baths, later a more
general approach was proposed which was based on the fact that memory
effects generically vanish exponentially \cite{makri1999-iterative}.
The idea was further developed to simulate the dynamics of a generic
nonequilibrium bias driven system \cite{weiss2008-iterative}.

Since only a finite memory length needs to be considered, the IF can
be truncated beyond a memory time \(\tau_c=N_s\delta t\) (here \(N_s\)
is an positive integer), which corresponds to the time beyond which
bath correlations can be ignored controllably. Therefore the total IF
can be written approximately as
\cite{makarov1994-path,makri1995-numerical,makri1999-iterative,segal2010-numerically,segal2011-nonequilibrium}
\begin{equation}
\begin{split}
\label{eq:IF-iterative}
     I(s_0^{\pm},\ldots,s_N^{\pm})
     &\approx I(s_0^{\pm},\ldots,s_{N_s}^{\pm})
       I_s(s_1^{\pm},\ldots,s_{N_s+1}^{\pm})\\
     &\quad\times\cdots I_s(s_{N-N_s}^{\pm},\ldots,s_N^{\pm})
\end{split}
\end{equation}
with
\begin{equation}
\label{eq:IFs}
     I_s(s_k^{\pm},\ldots,s_{k+N_s})=
     \frac{I(s_k^{\pm},\ldots,s_{k+N_s}^{\pm})}
     {I(s_k^{\pm},\ldots,s_{k+N_s-1}^{\pm})}.
\end{equation}
The approach becomes exact when \(\tau_c\to\infty\) and its physical
content is discussed in
Refs. \cite{segal2010-numerically,makri1999-iterative}.

To integrate Eq. \eqref{eq:IF-iterative} iteratively we define a
multiple time reduced density matrix
\(\tilde{\rho}_S(s_k^{\pm},\ldots,s_{k+N_s-1})\) with an initial value
\(\tilde{\rho}_S(s_0^{\pm},\ldots,s_{N_s-1}^{\pm})=1\); i.e., all of the
initial components are identity. Its first evolution step is dictated
by
\begin{equation}
     \tilde{\rho}_S(s_1^{\pm},\ldots,s_{N_s}^{\pm})=\int\dd{s}_0^{\pm}
     I(s_0^{\pm},\ldots,s_{N_s}^{\pm}),
\end{equation}
and beyond the first step the evolution step is given by
\begin{equation}
\begin{split}
     \tilde{\rho}_S(s_{k+1}^{\pm},\ldots,s_{k+N_s}^{\pm})
     &=\int\dd{s}_k^{\pm}\tilde{\rho}_S(s_k^{\pm},\ldots,s_{k+N_s-1}^{\pm})\\
     &\qquad\times I_s(s_k^{\pm},\ldots,s_{k+N_s}^{\pm}).
\end{split}
\end{equation}
Then the time-local (\(t_k=k\delta t\)) reduced density matrix is
obtained by summing over all intermediate states:
\begin{equation}
     \rho_S(t_k)=\int\dd{s}_{k-1}^{\pm}\cdots\dd{s}^{\pm}_{k-N_s+1}
     \tilde{\rho}_S(s^{\pm}_{k-N_s+1},\ldots,s_k^{\pm}).
\end{equation}

In practical calculation, we need to keep track of
\(\tilde{\rho}_S(s_{k+1}^{\pm},\cdots,s_{k+N_s}^{\pm})\) which is a \(2N_s\)
rank ``tensor.'' Suppose the size of Hilbert space of the system is
\(M\); then a space with size proportional to \(M^{2N_s}\) is needed
to store the tensor. Similarly, to store
\(I_s(s_{k}^{\pm},\ldots,s_{k+N_s}^{\pm})\) one needs a space with
size proportional to \(M^{2(N_s+1)}\). The space size increases
dramatically with increasing \(M\) and \(N_s\), which means for
practical calculations we need to ensure \(M\) and \(N_s\) are not too
large. In other words, the size of the system and truncation time
\(\tau_c\) need to be small; otherwise the system may lose feasibility
in numerical evaluation.

However, there is no restriction or difficulty in the development of
the method that the Hamiltonian must be time independent. On the
contrary, since it is an iterative method which is truncated in time
it is rather easy to deal with a time-dependent Hamiltonian. What we
need to do is just calculate the IF with the Hamiltonian in
corresponding time. Moreover, this method is a nonperturbative method
beyond the Markovian limit and thus it is well suited for the
investigation of long-time behaviors of driven systems.

\section{The Model}
\label{sec:model} 

In this article we consider the spin-fermion model where a spin is
coupled to two fermionic leads with chemical potential difference
\(\Delta\mu\) at zero temperature. Such a model serves as a simple but
nontrivial model to study bias driven nonequilibrium system
\cite{segal2007-nonequilibrium,segal2010-numerically,mitra2007-coulomb,lutchyn2008-quantum}.
The Hamiltonian of the model is written as
\begin{equation}
	H=H_0+H_1,
\end{equation}
where 
\begin{equation}
	H_0=H_S,\quad H_1=H_B+H_{SB}.
\end{equation}
The bath Hamiltonian \(H_B\) is that of two independent free fermion
baths (\(\alpha=L,R\)) whose statistics are determined by chemical
potentials, i.e.,
\begin{equation}
     H_B=\sum_{\alpha,k}\varepsilon_kc_{\alpha k}^{\dag}c_{\alpha k}.
\end{equation}
The operator \(c^{\dag}_{\alpha k}\) (\(c_{\alpha k}\)) creates
(annihilates) an electron with state \(k\) in the \(\alpha\)th
bath. The system Hamiltonian \(H_S\) is that of a driven TLS,
\begin{equation}
H_S=\frac{1}{2}\bm{B}(t)\cdot\bm{\sigma}
\end{equation}
where \(\bm{\sigma}=(\sigma_x,\sigma_y,\sigma_z)\) are Pauli matrices
and \(\bm{B}(t)\) is the external field. The system-bath coupling is
taken to be
\begin{equation}
\label{eq:Hsb}
H_{SB}=\sum_{\alpha\beta;kq}V_{\alpha\beta}
c_{\alpha k}^{\dag}c_{\beta q}\sigma_z,
\end{equation}
where \(\alpha,\beta=L,R\) are the bath indices. In this article we
focus on the model
\cite{segal2007-nonequilibrium,segal2010-numerically,ng1995-x,ng1996-fermi}
where the momentum dependence of the scattering potential is
neglected. In particular, we consider only interbath system-bath
couplings for which
\(gV_{\alpha\beta}=\lambda(1-\delta_{\alpha\beta})\), where \(g\) is
the density of states of each Fermi bath and \(\lambda\) is the
control parameter.

For numerical evaluation we need to employ a second-order
Trotter-Suzuki decomposition
\cite{trotter1959-product,suzuki1976-generalized} on the discrete
evolution operator \(e^{iH\delta t}\) for which
\begin{equation}
	e^{iH\delta t}\approx e^{iH_1\delta t/2}e^{iH_0\delta t}e^{iH_1\delta t/2}.
\end{equation}
With this decomposition and assuming separable initial conditions
\(\rho(0)=\rho_S(0)\rho_B(0)\), the IF of the present model can be
identified as
\begin{equation}
\begin{split}
	\label{eq:SF-IF}
	I(s_0^{\pm},\ldots,s_N^{\pm})&=\mel{s_0^+}{\rho_S(0)}{s_0^-}\\
	&\times K(s_N^{\pm},s_{N-1}^{\pm})\cdots K(s_1^{\pm},s_0^{\pm})\\
	&\times \Tr_B{}[e^{-iH_1(s_N^+)\delta t/2}e^{-iH_1(s_{N-1}^+)\delta t}\\
	&\qquad\times\cdots e^{-iH_1(s_1^+)\delta t}e^{-iH_1(s_0^+)\delta t/2}\\
	&\qquad\times\rho_B(0)e^{iH_1(s_0^-)\delta t/2}e^{iH_1(s_1^-)\delta t}\\
	&\qquad\times\cdots e^{iH_1(s_{N-1}^-)\delta t}e^{iH_1(s_N^-)\delta t/2}],
\end{split}
\end{equation}
where
\begin{equation}
\begin{split}
	K(s_{k+1}^{\pm},s_k^{\pm})&=\mel{s_{k+1}^+}{e^{-iH_0(t_k)\delta t}}{s_k^+}\\
		&\qquad\quad\times\mel{s_k^-}{e^{iH_0(t_k)\delta t}}{s_{k+1}^-}
\end{split}
\end{equation}
is the propagator matrix for the isolated system.

It is more flexible to describe the bath as discrete levels and the
infinite bath result can be easily reached even with a small number
(about 40) of effective bath fermions
\cite{segal2010-numerically}. The trace in Eq. \eqref{eq:SF-IF} can be
numerically eliminated via the Blankenbecler-Scalapino-Sugar (BSS)
identity \cite{blankenbecler1981-monte} and Levitov's formula
\cite{Klich2003-quantum,abanin2004-tunable,abanin2004-tunable,abanin2005-fermi};
then the analytic structure of the trace is not required. This feature
gives the method feasibility to investigate various system-bath
coupling other than linear coupling including nonadditive system-bath
coupling \cite{mu2017-qubit} used in our model. The generalization to
finite temperature is also straightforward
\cite{segal2010-numerically,Klich2003-quantum}.

\section{Floquet Formalism}
Alternatively, one could also use the Floquet master equation to study
time-dependent systems. Comparing to the iterative path integral
technique, the Floquet master equation approach is restricted to
periodically driven systems and it is based on perturbative
expansions.  We employ a Floquet Born-Markov master equation with
nonadditive system-bath interaction to calculate our model as a
benchmark.  We give a brief introduction to the Floquet formalism here
and a detailed derivation of the Floquet master equation is given in
Appendix \ref{sec:floquet-me}.

Let us consider an isolated system with Hamiltonian \(H_S(t)\). If the
driving field \(\bm{B}(t)\) is a periodic function with period
\(\mathcal{T}\) for which \(H_S(t)=H_S(t+\mathcal{T})\), then the
Floquet theorem states that \cite{shirley1965-solution}, for
Schr\"{o}dinger equation with system coordinate $q$,
\begin{equation}
i\pdv{t}\psi(q,t)=H_S(t)\psi(q,t)
\end{equation}
there exist solutions in the form
\begin{equation}
\psi_i(q,t)=e^{-i\varepsilon_i t}\varphi_i(q,t),
\end{equation}
where \(\varphi_i(q,t)\) is periodic in time with period
\(\mathcal{T}\) and \(\varepsilon_i\) is a real-valued function. The
term \(\varphi_i\) is called the Floquet state and term
\(\varepsilon_i\) is called the quasienergy. It is clear that
\(\varepsilon_i\) is unique up to multiples of
\(\Omega=2\pi/\mathcal{T}\) for which \(\varepsilon_i+n\Omega\) with
\(n\) being an integer corresponds to the same physical state. The
Floquet states \(\varphi_i(q,t)\) form a complete orthonormal basis
for the system at given time \(t\). The time evolution operator (for
\(t\ge t'\))
\begin{equation}
U_S(t,t')=\mathrm{T}\exp[-i\int_{t'}^t H_S(\tau)\dd{\tau}]
\end{equation}
then can be expressed in the Floquet basis by
\begin{equation}
U_S(t,t')=\sum_n\ket{\varphi_n(t)}\bra{\varphi_n(t')}e^{-i\varepsilon_n(t-t')}.
\end{equation}

In the Floquet basis, the density matrix can be defined as
\begin{equation}
\varrho_{ij}(t)=\mel{\varphi_i(t)}{\rho_S(t)}{\varphi_j(t)}.
\end{equation}
In this representation the time-dependent part of the Hamiltonian is
absorbed in the Floquet states \(\varphi_i(t)\); therefore, the
time-dependent part of \(\varrho_{ij}(t)\) is greatly simplified. In
particular, \(t\to\infty\) leads to the time-independent density
matrix element \(\varrho_{ij}\) which represents the asymptotic
Floquet states \cite{hone2009-statistical}. In Appendix
\ref{sec:floquet-me} we give a comparison between asymptotic Floquet
states calculated by the iterative path integral method used in this
article and those calculated by the Born-Markov Redfield master
equation in Floquet representation.

\section{Results}
\label{sec:results} 

Back in 1927, Hund \cite{hund1927-on} pointed out the importance of
the quantum tunneling effect in intramolecular rearrangements. Since
then quantum tunneling between two levels in isolated TLSs under
external driving has been widely studied. Rich tunneling phenomena are
found in such problems. For instance, several physicists, including
Landau, Zener, and St\"ueckelberg studied the transition between two
levels in isolated TLSs under a time-dependent energy sweep external
field
\cite{landau1932-on,zener1932-non,stueckelberg1932-theorie}. Such a
model is commonly known as Landau-Zener model and it has a wide range
of applications in physics and quantum chemistry. Another example is
the coherence destruction of the tunneling phenomenon for which in
isolated TLSs under monochromatic driving
\cite{grossmann1991-coherent} the tunneling could be suppressed by the
external driving.

According to our simulations, under monochromatic driving the
spin-fermion model would exhibit rich tunneling phenomena. Many of
them could also be found in the spin-boson model
\cite{grifoni1998-driven,makarov1995-control,magazzu2018-asymptotic}. These
may be common phenomena of driven dissipative TLSs regardless of what
kind of bath is present.  To shorten the length and reduce the number
of figures of this article, we do not represent them here and instead
focus on showing the results which are relevant to the robustness of
asymptotic behaviors of the system against different \(\lambda\). Both
circularly and linearly polarized driving fields are considered in
this article. Although tunneling behaviors of the system differ a lot
under different driving, under certain conditions all of them exhibit
robustness behaviors.

\subsection{Circularly Polarized Fields}
Let us first consider the case of a spin-fermion model driven by a
spatially homogeneous, circularly polarized field. We set the spin to
the \(z\) direction at the initial time \(t=0\), i.e.,
\begin{equation}
\label{eq:initial}
\expval{\sigma_z}(0)=1\quad\hbox{and}\quad\rho_S(0)=\mqty(1&0\\0&0\\).
\end{equation}

A pioneering work on isolated driven TLSs in a circularly polarized
field is given by Rabi \cite{rabi1937-space} and it is shown that in
this case analytic solutions can be found
\cite{rabi1937-space,grifoni1998-driven}. However, it is difficult to
find analytic solutions in general cases and thus for consistency we
simulated the isolated driven TLS numerically in this article.

\subsubsection{Field in \(x\)-\(y\) Plane}
\label{sec:circular-xy} 

Here we consider the case where the time-dependent field is orthogonal
to \(\expval{\sigma_z}(0)\), i.e., in the \(x\)-\(y\) plane. The
system Hamiltonian can be written as
\begin{equation}
H_S(t)=\frac{B}{2}\sigma_z+\frac{\Delta}{2}(\sigma_x\cos\Omega t+\sigma_y\sin\Omega t).
\end{equation}

Note that if we turn off the time-dependent part of the field, namely
set \(\Delta=0\) and only retain the static part
\(\frac{B}{2}\sigma_z\), then the system becomes localized; i.e.,
there would be no tunneling between two levels. In other words the
tunneling is totally induced by the time-dependent field and in the
dissipationless case \(\expval{\sigma_z}(t)\) would stay as 1 if
\(\Delta=0\) with the initial condition in Eq. \eqref{eq:initial}.

\begin{figure}
\centering
\subfigure{
     \includegraphics[scale=1]{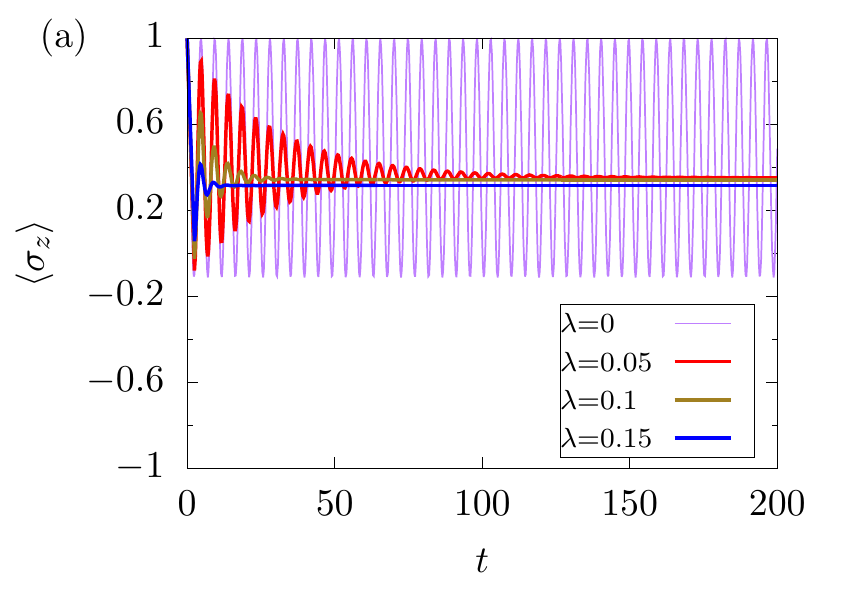}
     \label{fig:01}}
\subfigure{
     \includegraphics[scale=1]{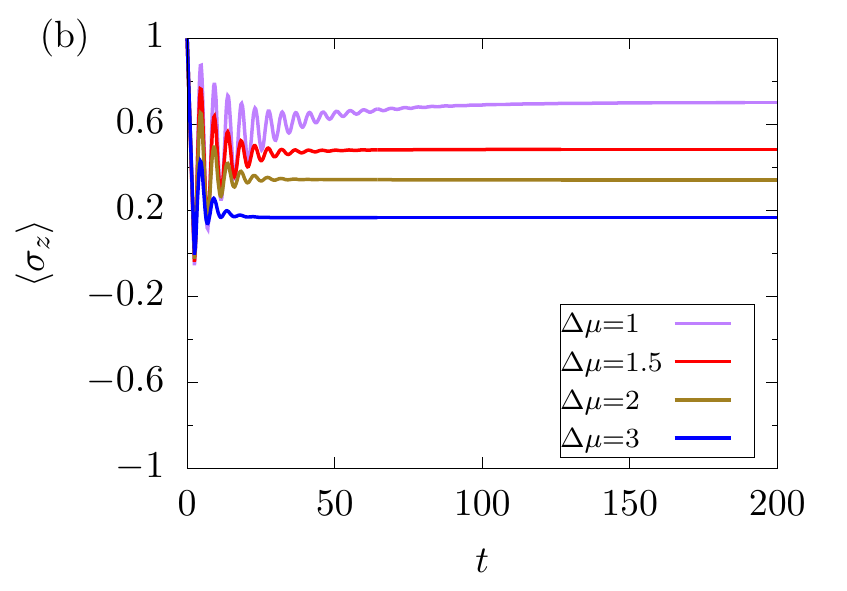}
     \label{fig:02}}
   \caption{(Color Online) Dynamics of $\expval{\sigma_z}(t)$ under a
     circularly driving field in the \(x\)-\(y\) plane for (a) varying
     system-bath coupling strength with $\Delta\mu=2$ and (b) varying
     chemical potential difference with $\lambda=0.1$.  We have used
     $B=0.1,\Delta=1$, and $\Omega=1$ for both (a) and (b).}
\label{fig:circular-xy}
\end{figure}

With small local potential \(B=0.1\) and under slow and weak driving
(\(\Delta=0.1,\Omega=0.1\)), the behavior of \(\expval{\sigma_z}(t)\)
is similar to that of a spin-fermion model without driving
\cite{segal2010-numerically}: the chemical potential difference acts
as a temperature like contributor to dephasing
\cite{segal2007-nonequilibrium} for which \(\expval{\sigma_z}(t)\)
would be eventually dissipated to zero. The dissipation rate would be
larger with larger \(\lambda\) and \(\Delta\mu\).

If the driving field is both fast and strong (\(\Delta=1\) and
\(\Omega=1\)) the driving and dissipation eventually reach a balanced
state and \(\expval{\sigma_z}(t)\) would remain a finite value instead
of decaying to zero, [see Fig. \ref{fig:01}]. At first glance this
behavior seems similar to the case in coherent destruction tunneling
\cite{grosmann1992-localization} (CDT) or driving-induced tunneling
oscillations
\cite{hartmann1998-dissipative,hartmann2000-driven,hausinger2010-dissipative}
(DITOs). However, the situation is different here. In CDT, without a
time-dependent field there would be tunneling between the two levels,
and the field suppresses the tunneling. Here the time-dependent field
induces, rather than suppresses, the tunneling between two levels. In
DITO, large amplitude oscillations are induced by a field with high
static energy bias \cite{hausinger2010-dissipative}, which is not in
our Hamiltonian. Without dissipation (\(\lambda=0\)),
\(\expval{\sigma_z}(t)\) oscillates in the positive region, which
means the system is tunneling between two levels but stays more time
in the spin-up state. Such tunneling is suppressed by the dissipation
and eventually the driving and dissipation reach a balanced state.

Here we can see an interesting phenomenon from Fig. \ref{fig:01}.
Lines have the same \(\Delta\mu\) but different \(\lambda\), and the
dissipation rate is larger with larger \(\lambda\). However,
\(\expval{\sigma_z}(t)\) with \(\lambda=0.05,0.1\) and 0.15 reach
almost the same value in asymptotic state, while from
Fig. \ref{fig:02} we could see that with the same \(\lambda\) but
different \(\Delta\mu\) the asymptotic \(\expval{\sigma_z}(t)\) differ
greatly. We could say that the system-bath coupling strength affects
the dissipation rate of the fast oscillation greatly but has little
effect on the asymptotic behavior, whereas the asymptotic behavior is
dominated by \(\Delta\mu\), or, in other words, by the electronic
structure of the baths. We may say the asymptotic Floquet states of
the system are \emph{robust} against different system-bath coupling
strengths \(\lambda\).  Such phenomena can be commonly seen in our
simulations and we discuss it further.

\subsubsection{Field in \(y\)-\(z\) Plane}
\label{sec:circular-yz}

Now consider the case where the time-dependent field is in the
\(y\)-\(z\) plane. The system Hamiltonian is
\begin{equation}
H_S(t)=\frac{B}{2}(\sigma_y\cos\Omega t+\sigma_z\sin\Omega t)
+\frac{\Delta}{2}\sigma_x.
\end{equation}

Note that in this case even if the time-dependent field is turned off,
namely, we set \(B=0\), the system is not localized since
\(\frac{\Delta}{2}\sigma_x\) remains in the Hamiltonian. The static
field can be viewed as a rotation axis along the \(x\) direction and
the spin would have uniform rotation around the axis. If we plot
\(\expval{\sigma_z}(t)\) in the dissipationless and static field case
we would see \(\expval{\sigma_z}(t)\) oscillates between 1 and \(-1\).

\begin{figure}
\centering
\subfigure{
     \includegraphics[scale=1]{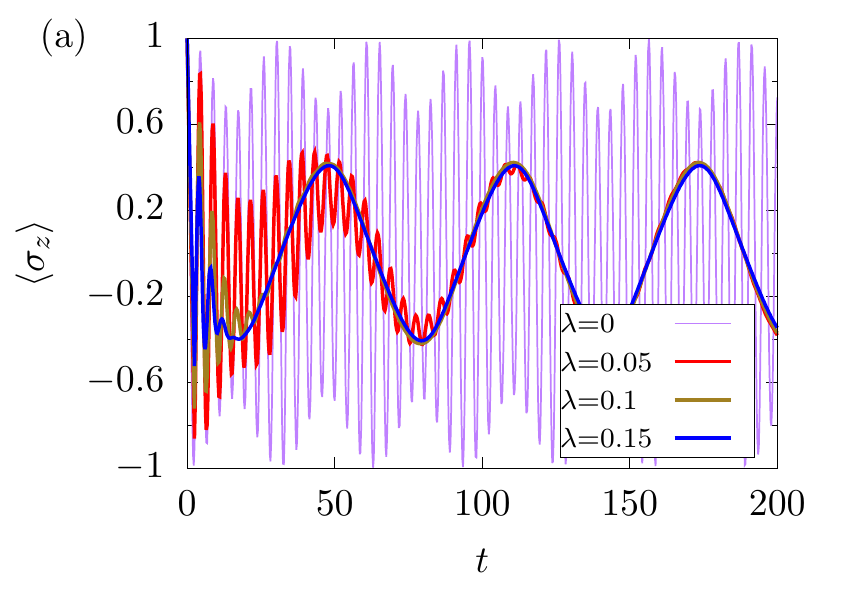}
     \label{fig:03}}
\subfigure{
     \includegraphics[scale=1]{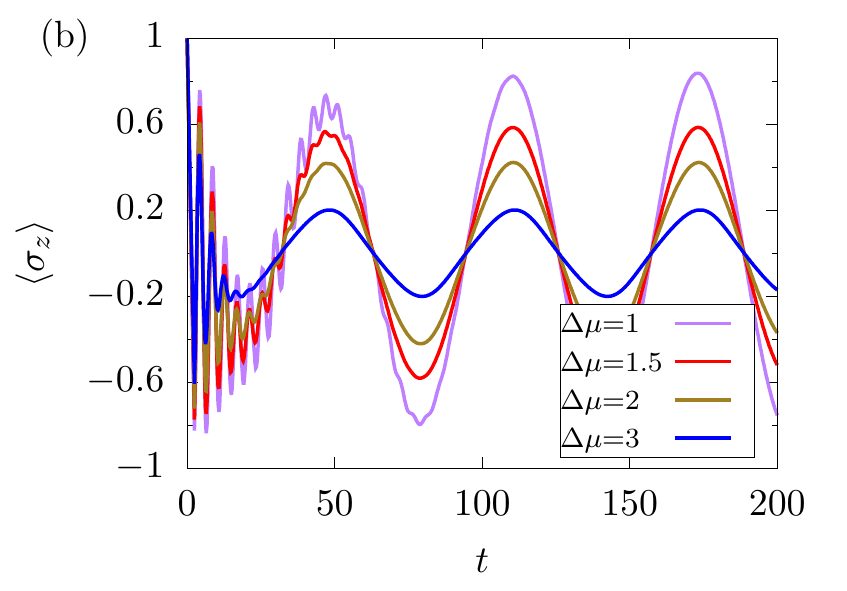}
     \label{fig:04}}
   \caption{(Color Online) Dynamics of $\expval{\sigma_z}(t)$ under a
     circularly driving field in the \(y\)-\(z\) plane for (a) varying
     system-bath coupling strength with $\Delta\mu=2$ and (b) varying
     chemical potential difference with $\lambda=0.1$.  We have used
     $B=1,\Delta=1$, and $\Omega=0.1$ for both (a) and (b).}
\label{fig:circular-yz}
\end{figure}

With large local potential \(\Delta=1\) and under strong but slow
driving (\(B=1,\Omega=0.1\)), coherent oscillations are induced (see
Fig. \ref{fig:circular-yz}). The fast oscillation part due to system
Hamiltonian dynamics is eventually dissipated out with a different
rate, and the coherent oscillation due to driving remains. Similar
behaviors have been reported in the spin-boson model
\cite{makarov1995-control,magazzu2018-asymptotic}. Such oscillation
behaviors present typical asymptotic Floquet states of a dissipative
spin system under periodic driving
\cite{makarov1995-control,magazzu2017-asymptotic,magazzu2018-asymptotic}.

Lines in Fig. \ref{fig:03} are simulated with the same \(\Delta\mu\)
but different \(\lambda\). It is worth noting that although their
dissipation rate is larger with larger \(\lambda\), the asymptotic
\(\expval{\sigma_z}(t)\) with different \(\lambda\) eventually
coincide.

Lines in Fig. \ref{fig:04} are simulated with the same \(\lambda\) but
different \(\Delta\mu\). It can be clearly seen that their final
\(\expval{\sigma_z}(t)\) differ greatly. Comparing Figs. \ref{fig:03}
and \ref{fig:04}, we may say that under this driving asymptotic
Floquet states of \(\expval{\sigma_z}(t)\) are dominated by the
electronic structure of the baths and they are robust against
different system-bath coupling strength \(\lambda\).

\subsection{Linearly Polarized Fields}
Let us next consider the case where a spin-fermion model is driven by
a spatially homogeneous, linearly polarized field. The initial
condition is the same as in Eq. \eqref{eq:initial}.  It was pointed
out long ago by Bloch and Siegert \cite{bloch1940-magnetic} that the
driven TLS problem is no longer analytically solvable when the field
is linearly rather than circularly polarized. To obtain an
approximating solution for a dissipationless driven TLS under a
linearly polarized field, the rotating wave approximation, which
approximately transforms the linearly polarized field to the form of a
circularly polarized field \cite{grifoni1998-driven}, is widely
used. An iterative approach for strong-coupling periodically driven
TLSs also exists \cite{wu2007-strong}. However, in this article we
directly use numerical results for the dissipationless case rather
than analytical approximations.

\subsubsection{Field in \(x\) Direction}
\label{sec:linear-x}

Here we consider the case where the time-dependent field is orthogonal
to \(\expval{\sigma_z}(0)\), say, along the \(x\) direction, and write
the system Hamiltonian as
\begin{equation}
H_S(t)=\frac{B}{2}\sigma_z+\frac{\Delta}{2}\sigma_x\cos\Omega t.
\end{equation}
With this Hamiltonian the system would be localized if the
time-dependent field is turned off. In other words, the tunneling
between two levels is totally induced by the time-dependent
field. This is a similar situation to that where the driving field is
circularly polarized in the \(x\)-\(y\) plane
(Sec. \ref{sec:circular-xy}).

\begin{figure}
\centering
\subfigure{
     \includegraphics[scale=1]{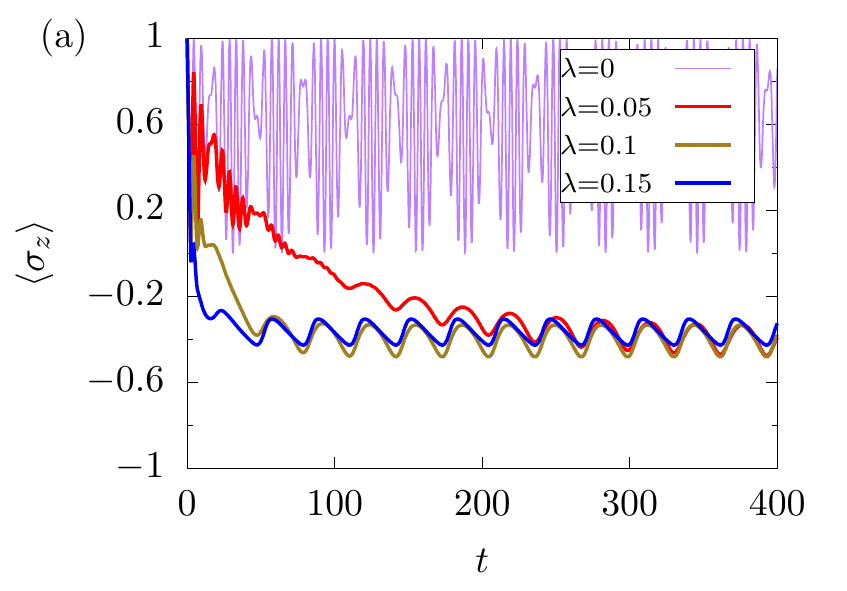}
     \label{fig:05}}
\subfigure{
     \includegraphics[scale=1]{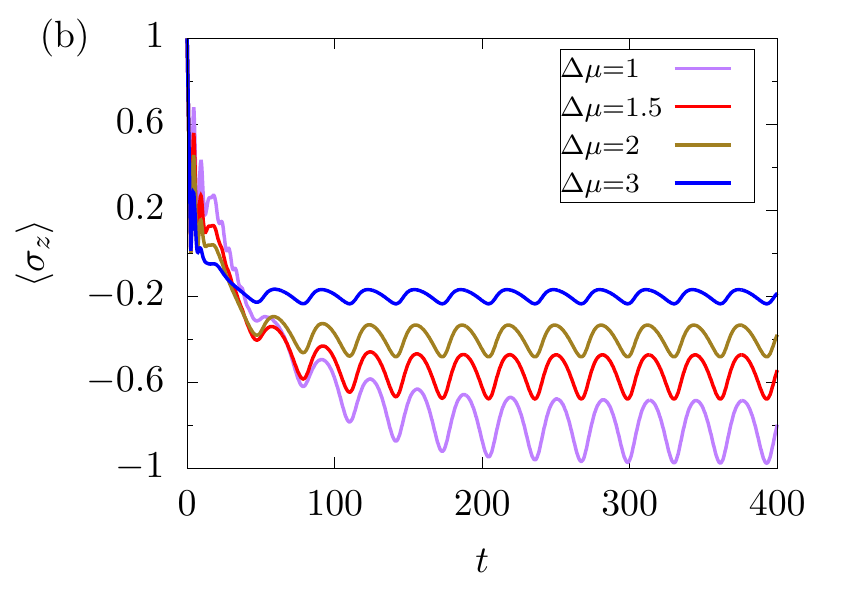}
     \label{fig:06}}
   \caption{(Color Online) Dynamics of $\expval{\sigma_z(t)}$ under a
     linearly driving field in the \(x\) direction for (a) varying
     system-bath coupling strength with $\Delta\mu=2$ and (b) varying
     chemical potential difference with $\lambda=0.1$.  We have used
     $B=1,\Delta=1$, and $\Omega=0.1$ for both (a) and (b).}
\label{fig:linear-x}
\end{figure}

In Fig. \ref{fig:linear-x}, the simulations are done under strong but
slow driving (\(\Delta=1\) and \(\Omega=0.1\)) and with large local
potential \(B=1\).

Figure \ref{fig:05} shows that \(\expval{\sigma_z}(t)\) oscillates
rapidly in the dissipationless (\(\lambda=0\)) case. When dissipation
is turned on, the system coherently oscillates around a nonzero value
when dissipation and driving are balanced. Such kinds of asymptotic
Floquet states are also reported in driven spin-boson models
\cite{magazzu2018-asymptotic}. The robustness occurs again for which
lines with same \(\Delta\mu\) but different nonzero \(\lambda\) almost
coincide eventually, whereas in Fig. \ref{fig:06} the lines have the
same \(\lambda\) but different \(\Delta\mu\), and they coherently
oscillate in different places.

\subsubsection{Field in \(z\) Direction}
\label{sec:linear-z}

Now let us consider the case where the time-dependent field is parallel
to \(\expval{\sigma_z}(0)\) with the system Hamiltonian
\begin{equation}
H_S=\frac{B}{2}\sigma_z\cos\Omega t+\frac{\Delta}{2}\sigma_x.
\end{equation}
There is always a static field along the \(x\) direction; thus, even
if the time-dependent driving field is off there is still tunneling
between the two levels. This is the usual case when studying the
dissipative TLS
\cite{makarov1994-path,makarov1995-control,makri1995-numerical,grifoni1995-cooperative,makri1997-stabilization,hartmann1998-dissipative,hartmann2000-driven,segal2007-nonequilibrium,segal2010-numerically,hausinger2010-dissipative,segal2014-heat,simine2013-path,magazzu2018-asymptotic};
thus, we present more simulations here.

\begin{figure}
\centering
\subfigure{
     \includegraphics[scale=1]{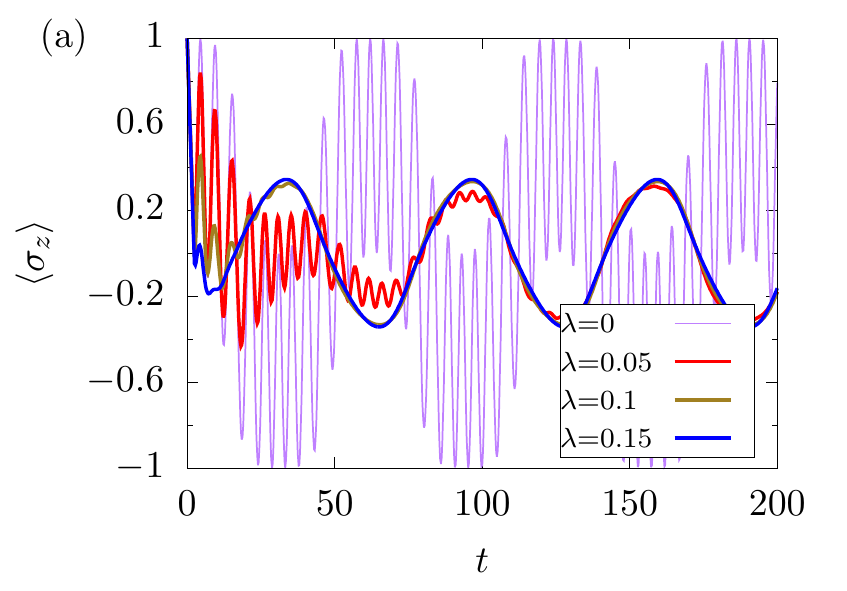}
     \label{fig:07}}
\subfigure{
     \includegraphics[scale=1]{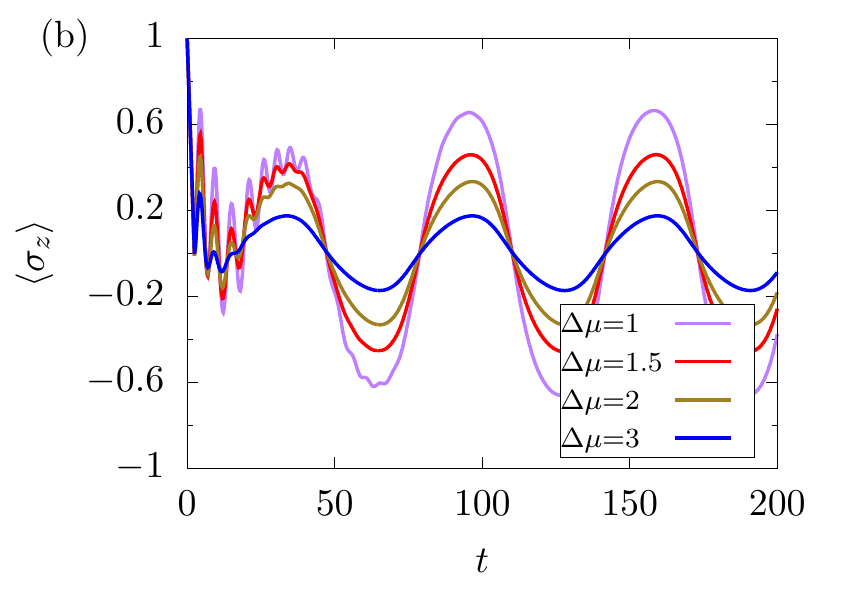}
     \label{fig:08}}
   \caption{(Color Online) Dynamics of $\expval{\sigma_z}(t)$ under a
     slow linearly driving field in the \(z\) direction for (a)
     varying system-bath strength with $\Delta\mu=2$ and (b) varying
     chemical potential difference with $\lambda=0.1$.  We have used
     $B=1,\Delta=1$, and $\Omega=0.1$ for (a) and (b).}
\label{fig:linear-z-slow}
\end{figure}

\begin{figure}
\centering
\subfigure{
     \includegraphics[scale=1]{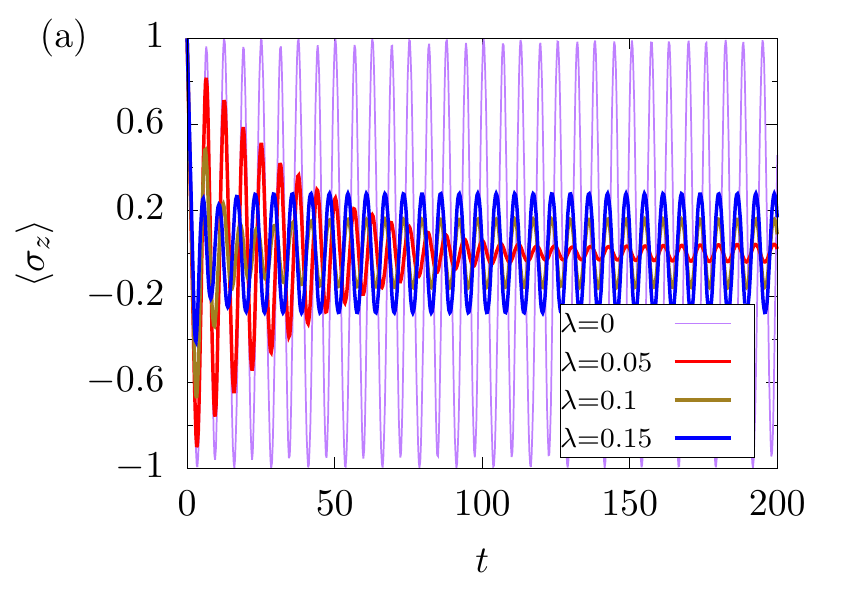}
     \label{fig:09}}
\subfigure{
     \includegraphics[scale=1]{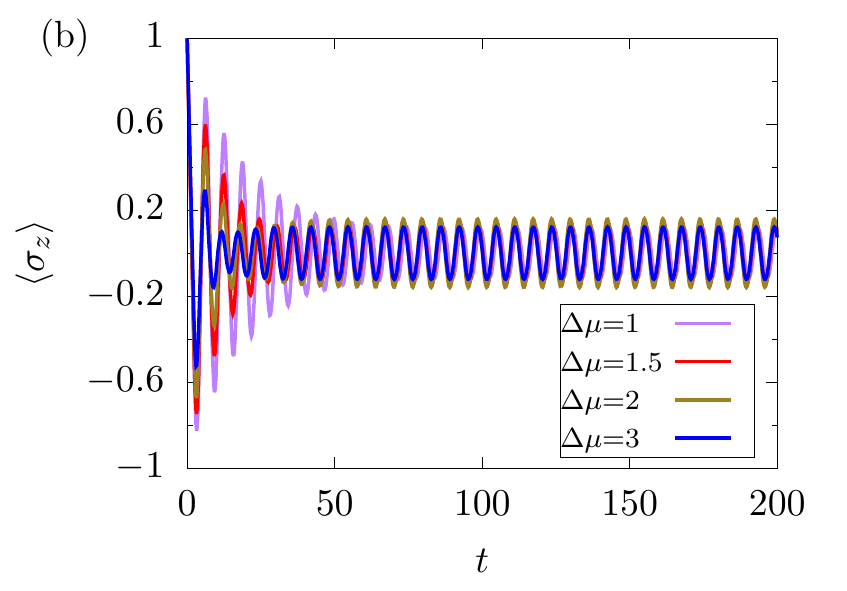}
     \label{fig:10}}
   \caption{(Color Online) Dynamics of $\expval{\sigma_z}(t)$ under a
     fast linearly driving field in the \(z\) direction for (a) varing
     system-bath coupling strength with $\Delta\mu=2$ and (b) varing
     chemical potential difference with $\lambda=0.1$.  We have used
     $B=1,\Delta=1$, and $\Omega=1$ for both (a) and (b).}
\label{fig:linear-z-fast}
\end{figure}

In Fig. \ref{fig:linear-z-slow} the simulations are done under strong
but slow driving (\(B=1,\Omega=0.1\)) with large local potential
\(\Delta=1\).  It can be seen under such driving that coherent
oscillations are induced which are similar to the circularly driving
field in the \(y\)-\(z\) plane case (Fig. \ref{fig:circular-yz}). It
can be seen from Fig. \ref{fig:07} that the asymptotic
\(\expval{\sigma_z}(t)\) with \(\Delta\mu=2\) and different
\(\lambda\) almost coincide. Under this driving the asymptotic Floquet
states exhibit robustness against different system-bath coupling,
whereas the asymptotic \(\expval{\sigma_z}(t)\) with \(\lambda=0.1\)
and different \(\Delta\mu\) differ a lot [see Fig. \ref{fig:08}].

The robustness \emph{breaks down} when we enter the fast driving
region by increasing \(\Omega\) to 1 [see Fig. \ref{fig:09}]. It can
be clearly seen that the asymptotic \(\expval{\sigma_z}(t)\) with
different \(\lambda\) differ a lot.  Interestingly, by intuition we
may think that the amplitude of oscillations with smaller \(\lambda\)
would be larger since dissipation is smaller. However, conversely, the
situation is that the amplitude of oscillations with the largest
\(\lambda=0.15\) is the largest while the amplitude of the line with
the smallest \(\lambda=0.05\) is the smallest. In other words,
stronger dissipation enhances the amplitude of oscillations of the
system rather than suppresses it. Such a phenomenon is a kind of
quantum stochastic resonance in which noises amplify and optimize the
response of a driven system
\cite{gammaitoni1998-stochasticresonance,makarov1995-stochastic,wagner2019-quantum}.

Figure \ref{fig:10} shows fast driving simulations with \(\lambda=0.1\)
and various \(\Delta\mu\). We found the situation is in another way
around, that the asymptotic \(\expval{\sigma_z}(t)\) coincide with
different \(\Delta\mu\).

Situations in Figs. \ref{fig:circular-yz} and \ref{fig:linear-z-slow}
are similar. However, if \(\Omega\) is increased to 1 in
Fig. \ref{fig:circular-yz} (not shown in this article) we could not
obtain similar results as in Fig. \ref{fig:10}: the robustness against
\(\lambda\) breaks down but the asymptotic \(\expval{\sigma_z}(t)\)
with different \(\Delta\mu\) would not coincide either.

At first glance we may conclude that the robustness against different
\(\lambda\) breaks down under fast driving. However, simulations in
Fig. \ref{fig:circular-xy} are also under fast driving but they still
show the robustness against \(\lambda\). Right now we could only propose
a hypothesis that the robustness would appear at least under slow
driving.

The results above indicate that there exists a complex interplay
between driving, bath electronic structure, and bath-system coupling.
Dissipation no longer simply acts as a contributor to dephasing under
time-dependent driving. Due to the interplay the robustness of
asymptotic Floquet states against \(\lambda\) is exhibited under
certain drivings. The interplay can lead to quantum stochastic
resonance as well.

\section{Discussion and Conclusions}
\label{sec:discussion}

We have numerically simulated the dynamics of the spin-fermion model
under various external monochromatic driving fields via a numerically
exact path integral method beyond the Markovian limit. Under
time-dependent driving, the spin-fermion model exhibits rich phenomena
which are not accessible in stationary situations.

We have also employed a Floquet master equation
\cite{hone2009-statistical} with the nonadditive \cite{mu2017-qubit}
system-bath interaction to check our results. The Floquet master
equation is in qualitative agreement with the path integral
method. However, the iterative path integral approach is
nonperturbative and can be applied to the system with nonperiodic
driving while the Floquet master equation, on the other hand, is
perturbative and restricted to periodic driving. To further validate
our approach, possible future work is to apply chain-mapping
approaches
\cite{vega2015-thermofield,guo2018-stable,tamascelli2018-efficient}
which consider the evolution of both system and bath and are also
nonperturbative.

It can be seen in our simulations that under a monochromatic driving
field coherent oscillation can be induced in many circumstances
(Figs. \ref{fig:circular-yz}, \ref{fig:linear-z-slow}, and
\ref{fig:linear-z-fast}). Such coherent oscillations are also reported
in the spin-boson model under a monochromatic linearly polarized
driving field in the \(z\) direction
\cite{makarov1995-control,magazzu2018-asymptotic}. Such oscillations
present typical asymptotic Floquet states of a dissipative spin system
under periodic driving
\cite{makarov1995-control,magazzu2017-asymptotic,magazzu2018-asymptotic}.

In addition, we also show that under strong and fast circularly
polarized driving field in the \(x\)-\(y\) plane
(Fig. \ref{fig:circular-xy}) the system exhibits behaviors similar to
CDT \cite{grosmann1992-localization} and DITO
\cite{hartmann1998-dissipative,hartmann2000-driven,hausinger2010-dissipative}. However,
as we pointed out in Sec. \ref{sec:circular-xy} there is an essential
difference between the behaviors in Fig. \ref{fig:circular-xy} and CDT
or DITO. In CDT or DITO, driving field suppresses rather than induces
the tunneling between the two states. Here \(\expval{\sigma_z}(t)\)
finally stays in a positive value because driving and dissipation
reach a balanced state.

A linearly polarized driving field in the \(x\) direction can induce
coherent oscillations around a nonzero value
(Fig. \ref{fig:linear-x}): without dissipation
\(\expval{\sigma_z}(t)\) oscillates around a nonzero value with wave
packets, while the dissipation suppresses fast oscillations and
\(\expval{\sigma_z}(t)\) eventually coherently oscillates around a
nonzero value.

Based on the observations of simulations in Sec. \ref{sec:results} we
find an interesting phenomenon that under certain drivings the
asymptotic Floquet states of the system exhibit a kind of robustness
against different system-bath coupling strength \(\lambda\). It can be
seen from Figs. \ref{fig:01}, \ref{fig:03}, \ref{fig:05}, and
\ref{fig:07} that in simulations with the same \(\Delta\mu\) but
different \(\lambda\) asymptotic \(\expval{\sigma_z}(t)\) almost
coincide. In other words, the asymptotic \(\expval{\sigma_z}(t)\) is
insensitive to \(\lambda\). Meanwhile another dissipative parameter
\(\Delta\mu\) would greatly affect the asymptotic
\(\expval{\sigma_z}(t)\). This shows that although both system-bath
coupling strength \(\lambda\) and chemical potential difference
\(\Delta\mu\) both contribute to dissipation, their effects are
distinct.

Such robustness breaks down when increasing \(\Omega\) to 1 in a
linearly driving field in the \(z\) direction case (see
Fig. \ref{fig:linear-z-fast}). The robustness also breaks down when
increasing \(\Omega\) to 1 in Fig. \ref{fig:circular-yz} (not shown in
this article). However, we cannot conclude directly that the
robustness breaks down under fast driving since simulations in
Fig. \ref{fig:circular-xy} (circularly polarized field in the
\(x\)-\(y\) plane) are also under fast driving (\(\Omega=1\)) but the
robustness still holds. Thus, for this moment we could only propose a
hypothesis that the robustness holds at least under slow driving.

In addition, in Fig. \ref{fig:09} quantum stochastic resonance is
demonstrated in which the amplitude of coherent oscillations is
enhanced, rather than suppressed, by stronger system-bath coupling
(stronger dissipation). Quantum stochastic resonance in the driven
spin-fermion model needs further theoretical and numerical
investigations and future study may be devoted to this issue.

These phenomena indicate that there exists a complex interplay between
the driving, bath electronic structure, and system-bath
coupling. According to our simulations the robustness against
different \(\lambda\) holds at least under slow driving, but what
plays the essential role may be ratios of all parameters and the form
of the driving field.

In conclusion, the spin-fermion model shows rich phenomena under
monochromatic driving. An interesting phenomenon is that under certain
drivings asymptotic Floquet states of the system exhibit robustness
against a range of system-bath coupling strength \(\lambda\): the
asymptotic behaviors of the system are insensitive to different
\(\lambda\) while the chemical potential difference of baths
\(\Delta\mu\) greatly affects them. Further simulations indicate that
the robustness may be essentially a result of the interplay between
the driving, bath electronic structure, and system-bath coupling and
thus can break down depending on the characteristics of the
interplay. The interplay can also lead to quantum stochastic
resonance.

The property of robustness indicates that under certain drivings the
asymptotic behaviors of the system are dominated by the electronic
structure of baths regardless of system-bath coupling strength. In
other words, we may extract information of the electronic structure of
baths without knowing the exact system-bath coupling strength. Such a
property may be useful in designing nanodetectors for the electronic
structure of the fermionic environment. Unlike the common sense
detector, the driving field is also a part of the detector and the
time-dependent parameters of the driving field are of essential
importance. Moreover, the observed result is also not merely a static
quantity but a time-dependent \(\expval{\sigma_z}(t)\). In this sense,
we may say it is a detector in time.

Open questions remain whether this property commonly exists in
dissipative systems and if the form of system-bath coupling changes or
the system is no longer a simple TLS.  Since parameters of numerical
simulations are limited by practical computation sources and
convergence conditions, theoretical analyses may give a more general
physical picture about the interplay between driving and
dissipation. Further numerical investigation on the driven spin-boson
model and theoretical investigation on the driven spin-fermion model
may be our future works.

\section*{Acknowledgements} We would like to thank Dvira Segal for the
discussion of implementations of the influence functional
technique. We also thank Jian-Sheng Wang and Hangbo Zhou for helpful
discussions.

\appendix
\section{Convergence and Error Analysis}
\label{sec:convergence}

There are three parameters relevant to accuracy and convergence of
simulations: bath size, the time step \(\delta t\), and the truncated
memory time \(\tau_c\). In this article, we have set the bath size as
80. With the identical fermionic bath, converged results are reported
for a bath size of 40 \cite{segal2010-numerically}. For a more
complicated system with both fermionic and bosonic baths, it has been
reported that bath size of 30 is sufficient to obtain converged
results \cite{simine2013-path}.

\begin{figure}
\centering
\subfigure{
     \includegraphics[scale=1]{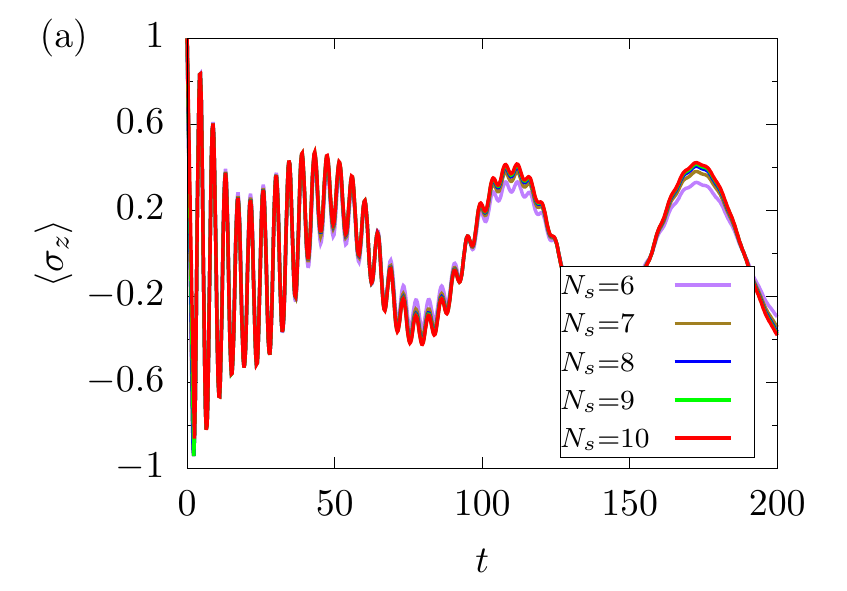}
     \label{fig:convz}}
\subfigure{
     \includegraphics[]{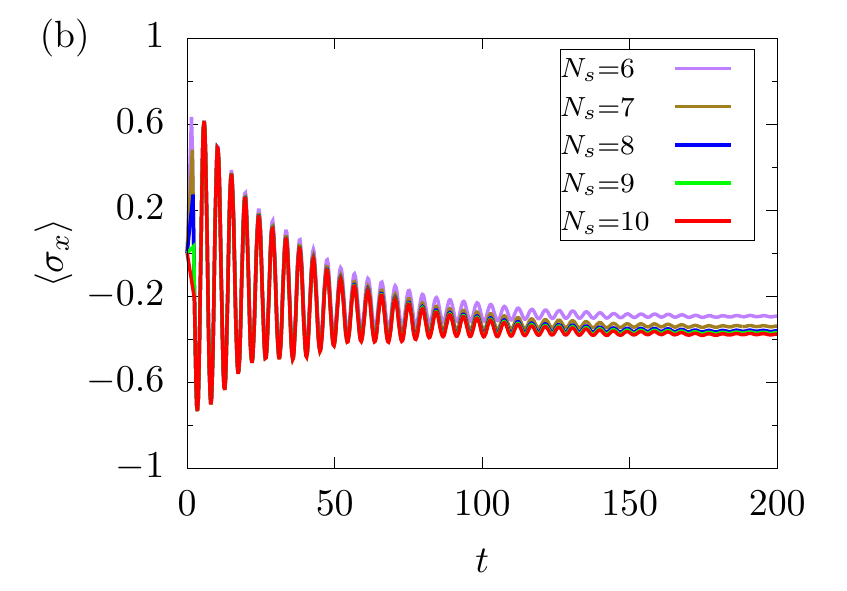}
     \label{fig:convx}}
   \caption{(Color Online) Convergence behavior with increasing
     $N_s$. It is a reexamination of Fig. \ref{fig:circular-yz} and
     other parameters are the same as those in
     Fig. \ref{fig:circular-yz}.}
\label{fig:convergence}
\end{figure}

Trotter error due to finite time step \(\delta t\) can be reduced by a
smaller \(\delta t\) and the result would become exact when
\(\delta t\to0\). However, with fixed \(N_s\) smaller \(\delta t\)
means smaller \(\tau_c\) and larger overall errors. Therefore, smaller
\(\delta t\) needs larger \(N_s\) to ensure enough length of
\(\tau_c\). As we discussed in Sec. \ref{sec:method}, the
computational memory increases exponentially with respect to \(N_s\)
and thus the value of \(N_s\) is also restricted according to the
available memory.

In principle one can extrapolate final results to the \(\delta t\to0\)
limit and the Trotter error is then eliminated
\cite{weiss2008-iterative,segal2010-numerically}. However, since we
are dealing with the time-dependent driving case, with different
\(\delta t\) the driving field is sampled in different time
grids. This brings an extra error in extrapolation. To extrapolate to
the \(t\to0\) case, we need to do simulations with a range of
\(\delta t\) with a fixed \(\tau_c\). When \(\delta t\) is large, the
time grid may miss important points, while small \(\delta t\) needs
large \(N_s\) which may be not acceptable in practical
computation. Therefore, in this article extrapolation is not employed.

It is shown that \cite{weiss2008-iterative,segal2010-numerically}
\(\tau_c\) roughly corresponds to \(1/\Delta\mu\) for the spin-fermion
model without driving. And \(\delta t=0.25\) and \(N_s=9\) are enough
to ensure convergence for small \(\Delta\mu=0.6\) and intermediate
coupling strength \(\lambda=0.2\). For strong interaction the time
step \(\delta t\) needs to be small to guarantee convergence and
correspondingly a large \(N_s\) is needed. That is, if
\(\delta t=0.1\) and \(\Delta\mu\sim0.6\) then an extensive
computation effort as \(N_s>16\) is required.

In this article we set \(\delta t=0.25\) and \(N_s=10\). Since for
strong interaction small \(\delta t\) is needed for small
\(\Delta\mu\) (otherwise numerical divergence may be encountered), for
safety we keep \(\Delta\mu\ge1\) and \(\lambda\) in the weak and
intermediate interaction region (\(\lambda=0.05,0.1\), and \(0.15\))
in this article.

Figure \ref{fig:convergence} shows a convergence test for increasing
\(N_s\) with parameters used in Sec. \ref{sec:circular-yz}. Basically
it is a reexamination of Fig. \ref{fig:circular-yz} under circularly
polarized fields in the \(y\)-\(z\) plane with
\(B=1,\Delta=1,\Omega=0.1,\lambda=0.1\), and \(\Delta\mu=2\). In
Fig. \ref{fig:convz} the dynamics of \(\expval{\sigma_z}(t)\) is
shown.  Besides, the dynamics of \(\expval{\sigma_x}(t)\) is also
demonstrated in Fig. \ref{fig:convx}. It can be seen that convergence
is well reached at \(N_s=10\).

\section{Comparison to Steady-State Solution of Master Equation in Floquet States Representation}
\label{sec:floquet-me}

For comparison we employ a Born-Markov master equation in Floquet
states representation to study the case in Sec. \ref{sec:circular-xy}
where the driving field is circularly polarized in the \(x\)-\(y\)
plane. In this case, the Floquet states of the isolated driven TLS can
be analytically solved \cite{grifoni1998-driven,rabi1937-space} such
that the Floquet states and corresponding quasienergies are
\begin{equation}
\label{eq:floquet-states}
\begin{cases}
\displaystyle\varphi_1(t)=\sqrt{\frac{\omega-\Lambda}{2\omega}}
\mqty(e^{-i\Omega t}\\\frac{\Delta}{\omega-\Lambda}),&
\varepsilon_1=\frac{1}{2}(\omega-\Omega);\\
\displaystyle\varphi_2(t)=\sqrt{\frac{\omega+\Lambda}{2\omega}}
\mqty(1\\-\frac{\Delta}{\omega+\Lambda}e^{i\Omega t}),&
\varepsilon_2=-\frac{1}{2}(\omega-\Omega),\\
\end{cases}
\end{equation}
where \(\Lambda=\Omega-B\) and \(\omega=\sqrt{\Delta^2+\Lambda^2}\).

Here following the derivation in
Ref. \cite{hone2009-statistical} we give a brief review about
the Floquet master equation technique. For conciseness, we denote the
system reduced density matrix \(\rho_S(t)\) by \(\varrho(t)\). After a
standard Born-Markov master equation procedure \cite{breuer2007-the},
we obtain an integro differential equation for \(\varrho(t)\),
\begin{equation}
\begin{split} 
\pdv{\varrho(t)}{t}=&-i[H_S(t),\varrho(t)]\\
&-\int_0^t[\sigma_z,\sigma_z(t-\tau,t)\varrho(t)]C(\tau)\dd{\tau}\\
&+\int_0^t[\sigma_z,\varrho(t)\sigma_z(t-\tau,t)]C^*(\tau)\dd{\tau},
\end{split}
\end{equation}
where \(\sigma_z(t-\tau,t)\) stands for
\(U_S(t,t-\tau)\sigma_zU_S^{\dag}(t,t-\tau)\).

Unlike most of the existing works based on two-body system-bath
interaction, our system-bath coupling given by Eq. \eqref{eq:Hsb}
contains nonadditive interactions. Therefore, our bath correlation
function \(C(\tau)\) is
\begin{equation}
\begin{split}
C(\tau)=&\sum_{\alpha\beta;\alpha'\beta'}\sum_{kq;k'q'}
V_{\alpha\beta}V_{\alpha'\beta'}
\langle c_{\alpha k}^{\dag}(\tau)c_{\beta q}(t)c_{\alpha'k'}^{\dag}c_{\beta'q'}\rangle\\
=&\frac{\lambda^2}{g^2}\sum_{\alpha\ne\beta}\sum_{kq}
n_{\alpha k}(1-n_{\beta q})e^{i(\varepsilon_k-\varepsilon_q)\tau},
\end{split}
\end{equation}
where \(g\) is the density of states of each Fermi bath and \(n_{\alpha
k}=\langle c_{\alpha k}^{\dag}c_{\alpha k}\rangle=\Tr_B[c_{\alpha
k}^{\dag}c_{\alpha k}]\). It is convenient to define the quantity
\(C(E)\) as
\begin{equation}
\begin{split}
C(E)=&\int_0^{\infty}C(\tau)e^{iE\tau}\dd{\tau}\\
\approx&\frac{\lambda^2}{g^2}\sum_{\alpha\ne\beta}\sum_{kq}
n_{\alpha k}(1-n_{\beta q})\delta(\varepsilon_k+E-\varepsilon_q),
\end{split}
\end{equation}
where principal value contributions are neglected. Note that a
continuous, not discretized, energy spectrum is used here to calculate
\(C(E)\).

Now we define the reduced density matrix of the system in Floquet
representation as
\(\varrho_{ij}(t)=\mel{\varphi_i(t)}{\varrho(t)}{\varphi_j(t)}\).
Denoting \(\mel{\varphi_i(t)}{\sigma_z}{\varphi_j(t)}\) by
\(\sigma_{ij}(t)\) and expanding it in Fourier series such that
\begin{equation}
\sigma_{ij}(t)=\sum_m\sigma_{ij}(m)e^{im\Omega t}
\end{equation}
yields the master equation for \(\varrho_{ij}(t)\),
\begin{equation}
\begin{split}
\qty(\pdv{t}+i\varepsilon_{ij})\varrho_{ij}(t)=&
-\sum_{kl}[R_{ik;kl}(t)\varrho_{lj}(t)-R_{lj;ik}(t)\varrho_{kl}(t)\\
&\quad-R^*_{ki;jl}(t)\varrho_{kl}(t)+R^*_{jl;lk}(t)\varrho_{ik}(t)],
\end{split}
\end{equation}
where [noticing that \(\sigma_{ij}^{*}(m)=\sigma_{ji}(-m)\)]
\begin{equation}
\begin{split}
R_{ij;kl}(t)=&\sum_{mn}e^{i(m+n)\Omega t}\sigma_{ij}(m)\sigma_{kl}(n)\\
&\qquad\times\int_0^t e^{-i(\varepsilon_{kl}+n\Omega)\tau}C(\tau)\dd{\tau},
\end{split}
\end{equation}
and
\begin{equation}
\begin{split}
R^*_{ij;kl}(t)=&\sum_{mn}e^{i(m+n)\Omega t}\sigma_{ji}(m)\sigma_{lk}(n)\\
&\qquad\times\int_0^t e^{-i(\varepsilon_{lk}+n\Omega)\tau}C^*(\tau)\dd{\tau}
\end{split}
\end{equation}
with \(\varepsilon_{ij}=\varepsilon_i-\varepsilon_j\).

\begin{figure}
\centerline{\includegraphics[]{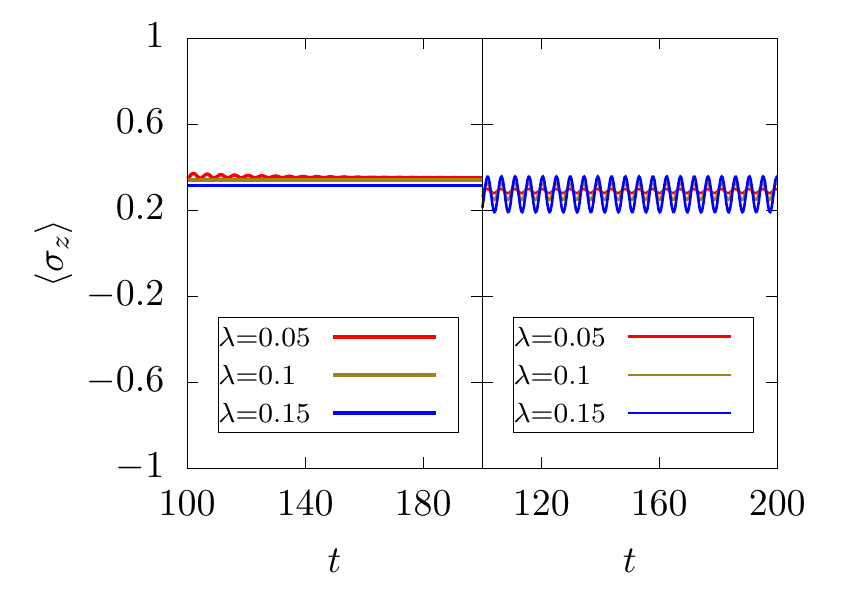}}
\caption{(Color Online) $\expval{\sigma_z}(t)$ in asymptotic Floquet
  states with different $\lambda$ calculated by different
  methods. Left: Results of the path integral method. Right: Results
  of the Floquet master equation. Other parameters are the same as
  those in Fig. \ref{fig:01}.}
\label{fig:floquet-lambda}
\end{figure}

\begin{figure}
\centerline{\includegraphics[]{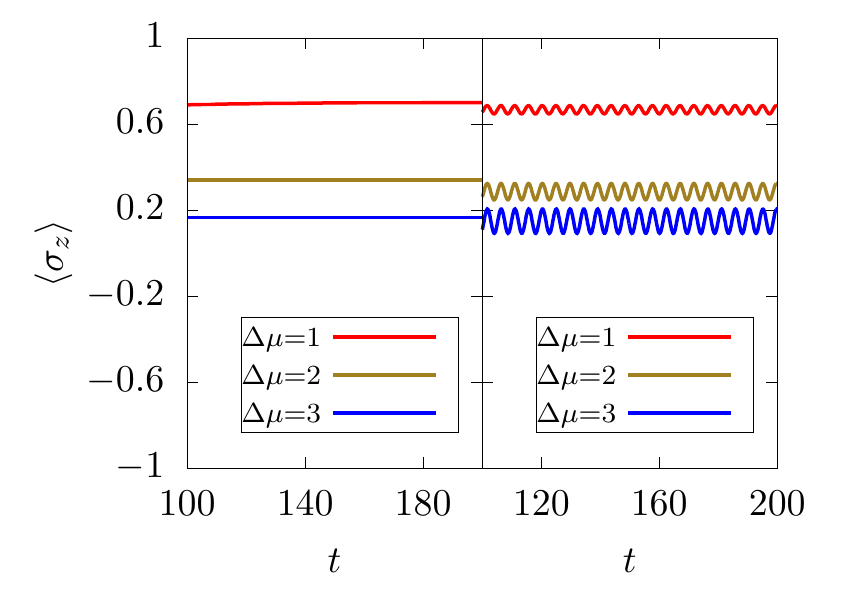}}
\caption{(Color Online) $\expval{\sigma_z}(t)$ in asymptotic Floquet
  states with different $\Delta\mu$ calculated by different
  methods. Left: Results of the path integral method. Right: Results
  of the Floquet master equation. Other parameters are the same as
  those in Fig. \ref{fig:02}.}
\label{fig:floquet-mu}
\end{figure}

In the steady states where \(t\to\infty\), only terms satisfying
\(m+n=0\) survive due to the vanishing of oscillating factors
\(e^{i(m+n)\Omega t}\). The master equation for steady states
\(\varrho_{ij}\) then reads
\begin{equation}
\label{eq:floquet-steady}
i\varepsilon_{ij}\varrho_{ij}=
-\sum_{kl}[R_{ik;kl}\varrho_{lj}-R_{lj;ik}\varrho_{kl}
-R^*_{kl;jl}\varrho_{kl}+R^*_{jl;lk}\varrho_{ik}],
\end{equation}
where
\begin{equation}
R_{ij;kl}=\sum_m\sigma_{ij}(m)\sigma_{kl}(-m)\int_0^{\infty}
e^{-i(\varepsilon_{kl}-m\Omega)\tau}C(\tau)\dd{\tau},
\end{equation}
and
\begin{equation}
R_{ij;kl}^*=\sum_m\sigma_{ji}(m)\sigma_{lk}(-m)\int_0^{\infty}
e^{-i(\varepsilon_{lk}-m\Omega)\tau}C^*(\tau)\dd{\tau}.
\end{equation}
According to Eq. \eqref{eq:floquet-states} the matrix element
\(\sigma_{ij}(t)\)'s are 
\begin{equation}
\begin{cases}
\sigma_{11}(t)=-\frac{\Lambda}{\omega},&
\sigma_{12}(t)=\frac{\Delta}{\omega}e^{i\Omega t},\\
\sigma_{21}(t)=\frac{\Delta}{\omega}e^{-i\Omega t},&
\sigma_{22}(t)=\frac{\Lambda}{\omega}.
\end{cases}
\end{equation}
Therefore, only following \(\sigma_{ij}(m)\)'s are nonzero:
\begin{equation}
\begin{cases}
\sigma_{11}(0)=-\frac{\Lambda}{\omega},&
\sigma_{12}(1)=\frac{\Delta}{\omega},\\
\sigma_{21}(-1)=\frac{\Delta}{\omega},&
\sigma_{22}(0)=\frac{\Lambda}{\omega}.
\end{cases}
\end{equation}
Accordingly only six \(R_{ij;kl}\)'s are nonzero: \(R_{11;11}\),
\(R_{11;22}\), \(R_{22;11}\), \(R_{22;22}\), \(R_{12;21}\), and
\(R_{21;12}\). The asymptotic Floquet states can be found by solving
Eq. \eqref{eq:floquet-steady}. The comparisons between
$\expval{\sigma_z}(t)$ in asymptotic Floquet states calculated by the
path integral method and by by Floquet master equation are shown in
Fig. \ref{fig:floquet-lambda} and \ref{fig:floquet-mu}.

Figure \ref{fig:floquet-lambda} shows $\expval{\sigma_z}(t)$ in
asymptotic states with different \(\lambda\) when
\(\Delta\mu=2\). Most parameters are the same as those in
Fig. \ref{fig:01}: \(B=0.1\), \(\Delta=1\), and \(\Omega=1\). The
right-hand panel shows the results calculated by the Floquet master
equation.  It could be seen that although oscillation amplitudes of
\(\expval{\sigma_z}(t)\) are different with different \(\lambda\),
their mean position are almost the same. In other words, the Floquet
master equation calculation reproduces the robustness against
different \(\lambda\). The left-hand panel shows the results
calculated by the path integral method for comparison. It is shown
that the results of the two methods are in agreement except that the
results by Floquet master equation are weakly oscillating.

Figure \ref{fig:floquet-mu} shows $\expval{\sigma_z}(t)$ in asymptotic
states with different \(\Delta\mu\) when \(\lambda=0.1\). The mean
positions of $\expval{\sigma_z}(t)$ calculated by two methods are in
agreement. Both methods show that the results are significantly
altered by different chemical potential bias $\Delta\mu$.

The results in Fig. \ref{fig:floquet-lambda} and \ref{fig:floquet-mu}
show that the dynamics of \(\expval{\sigma_z}(t)\) by the path
integral method are in agreement with those of the Floquet master
equation. However, the Floquet master equation gives weakly
oscillating $\expval{\sigma_z}(t)$, whereas the oscillation decays to
zero when the path integral method is used. Since some other results
of the path integral method still show small oscillations (for
example, Fig. \ref{fig:linear-x}), the vanishing of the oscillation is
unlikely due to the numerical feature of the path integral method. The
origin of such oscillation by the master equation is possibly the
perturbative nature of the master equation. However, since these two
methods adopt different approximation schemes and numerical error
mechanisms, further analysis is needed to account for their numerical
difference.

\bibliographystyle{apsrev4-1}
\end{document}